# Statistical Models for the Number of Successful Cyber Intrusions

Nandi O. Leslie, Richard E. Harang, Lawrence P. Knachel, and Alexander Kott

U.S. Army Research Laboratory, Adelphi Laboratory Center

2800 Powder Mill Rd

Adelphi, MD 20783, USA

Email: Nandi.O.Leslie.Ctr@mail.mil

## Abstract

We propose several generalized linear models (GLMs) to predict the number of successful cyber intrusions (or "intrusions") into an organization's computer network, where the rate at which intrusions occur is a function of the following observable characteristics of the organization: (i) domain name server (DNS) traffic classified by their top-level domains (TLDs); (ii) the number of network security policy violations; and (iii) a set of predictors that we collectively call "cyber footprint" that is comprised of the number of hosts on the organization's network, the organization's similarity to educational institution behavior (SEIB), and its number of records on scholar.google.com (ROSG).  In addition, we evaluate the number of intrusions to determine whether these events follow a Poisson or negative binomial (NB) probability distribution.  We reveal that the NB GLM provides the best fit model for the observed count data, number of intrusions per organization, because the NB model allows the variance of the count data to exceed the mean.  We also show that there are restricted and simpler NB regression models that omit selected predictors and improve the goodness-of-fit of the NB GLM for the observed data.  With our model simulations, we identify certain TLDs in the DNS traffic as having significant impact on the number of intrusions. In addition, we use the models and regression results to conclude that the number of network security policy violations are consistently predictive of the number of intrusions.

## 1   Introduction

Our research aims to develop statistical models that predict the number of successful cyber intrusions into an organization's computer network, as a function of readily observable characteristics of the organization.  Whether such a model is even possible (i.e., whether the phenomena of cyber intrusions exhibit sufficient regularity for such a purpose) is in and of itself a research question we seek to answer.

We use the term "successful cyber intrusion" to describe a condition where a malware has been deposited on the computing devices of an organization, and an activity of the malware has been detected and confirmed by qualified personnel (also referred to as cyber analysts, in this paper) responsible for defense of the organization's computers and networks.  Merely an alert, or multiple alerts, such as provided by intrusion detection and prevention systems (IDPS), does not necessarily constitute a successful intrusion (Roesch, 1999).  Such alerts might constitute evidence of a successful intrusion, but often they do not.  Furthermore, a successful intrusion does not necessarily indicate a loss of data, or any other negative impact on the organization's data and computers assets. The malware may exhibit observable activity before it inflicts any damage, or the damage may be unknown. For the sake of brevity, in the remainder of this paper, we simply refer to intrusions as opposed to successful cyber intrusions.

Among organizations, the numbers of intrusions differ dramatically by many orders of magnitude. Some organizations experience a large number of intrusions in a given time frame, whereas others may not experience any intrusions for a number of years. A model capable of predicting the number of intrusions would be of significant value to providers of cyber security and resilience services. A specialized organization, such as a Managed Security Service Provider (MSSP), is often used by an organization to provide cyber defense services (Ding and Yurcik, 2005). An MSSP monitors the organization's computer networks, analyzes the information obtained from the network, detects intrusions and activities of malware on the network, and reports such detections to the operators of the network. The operators then take measures necessary to recover from the intrusion (for a more detailed discussion of MSSP operations, see Ding and Yurcik, 2005).

For an MSSP, the costs of doing business are heavily influenced by the number of intrusions experienced by its clients. Each suspected intrusion requires careful analysis by a well-qualified–and therefore relatively expensive–cyber analyst to minimize costly false-positive results. Once an intrusion is confirmed by analysis, the MSSP has to work with the client's system operators to find an appropriate way to (i) remedy the intrusion, (ii) minimize its impact, and (iii) formulate suitable measures to prevent such intrusions in the future. These are costly activities. Therefore, when a MSSP negotiates its fees with a new prospective client, it needs to estimate how many intrusions should be expected over some fixed time period. Past statistics for the given client are often unavailable, or can be seriously misleading. For example, if a less capable cyber defender served the client in the past, the number of detected intrusions could have been far lower than the actual number. Moreover, organizations can change rapidly, especially in ways that influence intrusion frequency; for example, an organization may change its business model, experience a merger, come under new management, enter new markets, etc. Therefore, an MSSP must devise a way to predict the potential number of future intrusions by means other than mere extrapolation of past experiences of the organization.

Another example where such a model would be of high value is its use for actuarial purposes. Highly damaging cyber-attacks, which are destructive in terms of both money and reputation, are routinely and widely discussed in the media, bringing questions of whether insurance protection is adequate against the risks of such incidents. The insurance industry has initiated significant efforts to research cyber risk, which is highly dependent on the probability of intrusions, and to develop insurance products for cyber risks. Such products are beginning to be seen as a necessity for doing business (Grande, 2014). Biener et al. (2014) describe the state of the insurance industry with respect to cyber risk and note its rapid growth as well as a number of problems in insurability of cyber risk. Modeling the probability of successful cyber intrusions, and collecting empirical data for validation of such models, are critical to resolving the insurability problems.

In a broader sense, a model of this nature contributes to our fundamental understanding of cyber situational awareness and ways to monitor, quantify, and manage cyber risk. The prominence of cyber risk has been rising in the last few years, partly due to the growing recognition of such risks to industries and governments. In particular, the U.S. Government has been pursuing a cybersecurity strategy that involves monitoring and assessing cyber risks and cyber situational awareness; such as monitoring for cyber intrusions, mitigation of vulnerabilities, improved authentication, and central collection of security incident data (Dempsey et al., 2011; DoD CIO, 2013). For example, as early as 2010, the Office of Management and Budget directed federal agencies to use automated continuous monitoring and risk models (Executive Office of the President, 2010). In 2012, the Department of Defense (DoD) released a document on its mobile device strategy that specifies continuous risk monitoring as part of the device management service (DoD CIO, 2012a). Later, it issued a document on cloud computing strategy that similarly gives a major role to risk monitoring (DoD CIO, 2012b). Furthermore, in 2013, the DoD identified improving cyber situational awareness through enhancing the cyber sensing infrastructure as a strategic imperative (DoD CIO, 2013). To this end, predictive models of cyber intrusions would be invaluable for improving cyber sensing infrastructure (e.g., intrusion detection and prevention systems) and address the difficult task of defending mission systems.

Models that can assist in automated quantification of risk, especially including the predicted number of intrusions, can be of great value (Kott and Arnold, 2013) as they open doors to comprehensive risk management decision-making. Employees at multiple levels—from senior leaders to system administrators—will be aware of continually updated risk distribution over the network components, and they will use this awareness to prioritize application of resources to the most effective remedial actions (Kott, 2014). Quantification of risks is not only important to improve the human decision-making process, but it can also contribute to rapid, automated or semi-automated implementation of remediation plans. Even more broadly, such quantitative models are critical tools in the emerging science of cyber security (Kott, 2014).

Finally, a model of this nature may offer clues towards enhancing the security posture and perhaps the design and operation of an organization's computing systems and networks. If the model indicates that certain characteristics are associated with increased number of intrusions, the organization might be able to find ways to modify those characteristics.

The remainder of this paper is organized as follows: in Section 2, we present additional related work; in Section 3, we describe the cyber intrusions data provided by an MSSP for analysis; in Section 4, we present a description for several generalized linear models (GLMs) for count data; in Section 5, we present our statistical analyses and findings; in Section 6, we discuss our conclusions.

## 2 Related work

There is surprisingly little prior research directly relevant to our objectives (i.e., modeling the number of intrusions as a function of an organization's characteristics). However, recently, there has been some interest in cyber-attack prediction with statistical and computational modeling techniques; such as attack graphs, Dynamic Bayesian Network (DBN), hidden Markov Model for time series, and Variable Length Markov Model (VLMM) (Yang et al., 2014; Gandotra et al., 2015; Zhan et al., 2015). Each of these modeling approaches requires a detailed understanding of the network and system vulnerabilities and the adversary's attack behavior (or attack plans) for implementation (Yang et al., 2014; Gandotra et al., 2015; Zhan et al., 2015). Our work differs from these approaches in that we analyze a combination of factors related to network traffic and organizational and user behaviors that have not been previously studied, and we show which of these factors are predictive of cyber intrusions.

In addition, there are several bodies of literature that are tangentially related to our work. For example, industry surveys (PWC, 2016) issued annually or more frequently, usually claim continuing and rapid growth in the overall volume of cyber-attacks, as well as the growth in size of data breaches. However, even the direction of trends in volume of cyber-attacks seem to be in dispute. For example, Edwards et al. (2015) analyze the dataset of data breaches (P. R. Clearinghouse, 2016) and find that neither size nor frequency of data breaches has increased over the past decade.

Another relevant line of research involves the collection of subjective assessments of an organization's policies and practices. This is usually done by sending questionnaires to Information Technology (IT) managers of multiple organizations. Although these results can shed light on correlations between management practices and subjective perceptions of security, such research does not extend into hard data regarding predicting the number of intrusions per organization (Khidzir et al., 2010; Hall et al., 2011; Yildirim et al., 2011; Fachkha et al., 2013), and quantitative modeling of the number of intrusions is not a feature of such research.

An additional related body of literature focuses on approaches to quantifying risk (e.g., so-called risk scoring algorithms). These are often limited to ad hoc heuristics, such as simple sums of vulnerability scores or counts of things (e.g., missing patches, open ports). Weaknesses and the potentially misleading nature of such metrics have been pointed out by a number of authors (Bartol et al., 2009; Jansen, 2009). For example, the individual vulnerability scores are dangerously reliant on subjective and qualitative input, all of which are potentially inaccurate and expensive to obtain. The presence of vulnerabilities is often unknown and depend on characteristics of computer hosts in a complex manner (Gil, Kott, and Barabasi, 2014). Similarly, neither network topology nor the roles and dynamics of inter-host interactions are considered by simple sums of vulnerabilities or missing patches. In general, there is a pronounced lack of rigorous theory and models of how various factors might combine into quantitative characterization of true risks, although there are efforts (Lippman et al., 2012) to formulate scientifically rigorous methods of calculating risks. This area of research does not attempt to use empirical data related to the number of intrusions, and it does not model such numbers as a function of characteristics of the organizations that are victims of intrusions.

A few research teams explore how certain statistical characteristics of organizations relate to those of intrusions. For example, using a database of data losses (https://blog.datalossdb.org/), Maillart and Sornette (2010) studied the probability distribution of ID losses per intrusion in the context of cyber risk, and they showed that the probability density functions are the same irrespective of the sectors of activity (e.g., academia, industry). Furthermore, a size effect is demonstrated, where the number of ID losses per event increases faster than linearly with organization size (Maillart and Sornette, 2010). Here, however, we show that organization type significantly impacts the number of intrusions; whereas, organization size as measured by the number of hosts is not a statistically significant predictor of the number of intrusions.

Julisch (2013) presents a qualitative study of an organization's cyber footprint to improve cyber preparedness, where cyber footprint, in that study, refers to the release of relevant security information by users on their public profiles (e.g., Facebook, Skype). In our study, we expand the scope of cyber footprint to also cover the release of information that may be indirectly relevant to the organization's cybersecurity (e.g., research publications, patents). We hypothesize that cyber footprint along with other features (presented in Section 3) comprise a set of yet untested predictors for intrusions, and we model this novel association to reveal how cyber footprint influences the number of intrusions. With our simulations, we investigate how these features and their differences among organizations influence the predicted number of intrusions.

In addition, there is a strong statistical relationship between network mismanagement and misuse symptoms (e.g., open recursive DNS resolvers, HTTPS server certificates, BGP misconfiguration) and cyber-attacks, by analyzing the correlation coefficients between mismanagement and maliciousness (Zhang et al., 2014). Moreover, there is research evaluating network mismanagement symptoms (e.g., DNS or BGP misconfiguration) in real-time to detect anomalous or illegal cyber behaviors and enhance IDPS with machine learning techniques, including principal component analysis (PCA), K-means, and Random Forest algorithms (Camacho et al., 2014; Cui et al., 2014; Liu et al., 2015). Our study differs from these studies in that we model the number of intrusions instead of modeling intrusion detection. Given the relationship between network mismanagement and misuse symptoms and intrusions presented in previous studies (Camacho et al., 2014; Cui et al., 2014; Zhang et al., 2014; Liu et al., 2015), we examine network security policy violations as a symptom of network misusage and hypothesize that it significantly impacts the number of intrusions.

In general, the lack of suitable empirical data is well recognized in the research literature. A paper by Cherdantseva et al. (2016) systematically examines 24 risk assessment methods. As far as we are able to determine, none of the methods involve actual numbers of intrusions or correlations of such numbers with other properties of the organizations or systems. The authors themselves note that lack of objective data hinders the validation of methods. Similarly, Biener et al. (2014) write, "There is a great need for more research on cyber insurance. Lack of data is a problem, however… Modelling cyber risk holds a great deal of promise, especially if data become available against which to test the models." In this light, our study is rather unique: we were able to obtain an adequate set of cyber intrusions data per organization, where each observation provides us with the number of intrusions along with a set of quantitative and categorical characteristics of each organization. Furthermore, we predict the number of cyber-attacks with GLMs implemented with jackknife (or leave-one-out) cross-validation for our regression results (see Sections 4 and 5 for a more detailed discussion). Therefore, the predictors for the models contribute distinctive insights into the drivers of cyber-attacks (see Table 1).

# 3 Data

We examine multiple data sources–provided to us by an MSSP–that together describe the number of intrusions for 41 client organizations along with other observed features for each organization: (i) security incident reports containing detailed information about malicious activities and computer security policy violations by users and operators; (ii) DNS traffic, collected with specialized and open source software for all organizations in our study; and (iii) other data sources describing a selected subset of features of each organization's network topology and cyber footprint (see further details below in Section 3.1). The definitions of the data are summarized in Table 1.

## 3.1 Predictor variables

In the subsections below, we categorize the predictors into three areas: (i) DNS traffic, (ii) security posture, and (iii) cyber footprint. In addition, we present descriptive statistics for the predictors in Table 1—these variables represent potential indicators of cyber intrusions. We test these hypotheses in subsequent sections to determine which of the potential indicators in Table 1 are in fact predictive of malicious activity.

### 3.1.1 DNS traffic

There are many recent intrusion detection studies that find features of DNS traffic to be predictive of malicious activities (Camacho et al., 2014; Cui et al., 2014; Zhang et al., 2014; Liu et al., 2015). In our study, we also hypothesize that DNS activities can help predict the number of intrusions that an organization may face. In addition, we observe and classify the domain names in the DNS traffic for 41 organizations by top level domain (TLD) and whether the TLDs resolve to US-based (domestic) or foreign IP addresses (see Table 1). The DNS data was captured using a combination of open source tools designed for network packet capture and in-house developed scripts to insert the data into a database. The capture was running 24 hours per day, and during this time, network activity for each cyber sensor was stored and analyzed, where there is typically one sensor per geographical site/organization monitored by the MSSP.

Table 1. Descriptive statistics for predictor variables by category.

| Predictors | Description | Min | Max | Mean | Median | Std dev |
|---|---|---|---|---|---|---|
| DNS traffic | | | | | | |
|   Domestic.com | Number of domestic.com visits | 12 | 1.3E+06 | 3.7E+05 | 2.8E+05 | 3.6E+05 |
|   Domestic.edu | Number of domestic.edu visits | 0 | 1.2E+04 | 2017.2 | 673.0 | 3203.4 |
|   Domestic.gov | Number of domestic.gov visits | 0 | 6358.0 | 1273.7 | 538.0 | 1770.6 |
|   Domestic.net | Number of domestic.net visits | 0 | 1.2E+06 | 1.6E+05 | 8.6E+04 | 2.3E+05 |
|   Domestic.org | Number of domestic.org visits | 0 | 5.5E+04 | 1.1E+04 | 4741.0 | 1.5E+04 |
|   Foreign.com | Number of foreign.com visits | 0 | 2.6E+05 | 4.5E+04 | 2.3E+04 | 5.7E+04 |
|   Foreign.net | Number of foreign.net visits | 0 | 1.3E+05 | 1.7E+04 | 7368.0 | 2.9E+04 |
|   Foreign.org | Number of foreign.org visits | 0 | 3.7E+06 | 4.2E+05 | 1.3E+05 | 6.8E+05 |
| Security posture | | | | | | |
|   violations | Number of policy violations | 0 | 24 | 5.1 | 4.0 | 5.1 |
| Cyber footprint | | | | | | |
|   hosts | Number of hosts | 15 | 3.2E+04 | 2145.9 | 500.0 | 5555.0 |
|   ROSG | Number of records on scholar.google.com (ROSG) | 1 | 1.8E+04 | 2753.8 | 312.0 | 4873.8 |
|   SEIB | Qualitative predictor representing an organization's similarity to academia and R&D institutions with 3 categories:<br>(i) Majority of organizations (78.0%) are neither academic nor R&D institutions and are assigned a 1 value<br>(ii) Research-oriented organizations are 17.1% of organizations with ROSG greater than 1,000 are assigned 3 value<br>(iii) Educational institutions make up 4.9% of organizations and are assigned 10 value | | | | | |

Specifically, we characterize the DNS traffic data for each organization by the number of foreign and domestic TLDs visited: .com, .org, and .net domains that resolved to a foreign-based IP address; and .edu, .gov, .org, .com, and .net that resolved to a US-based IP address (see Table 1). There are other TLDs that were captured in the data collection that were discarded (e.g., .co.uk, .bank, .science) so that we focus our initial analysis on the more popular TLDs (see Table 1). This method of categorizing the DNS data is designed to help us determine whether user internet behavior impacts the number of cyber-attacks.

### 3.1.2 Security posture

Intrusions are potentially more frequent in those networks, where the users and operators of the networks exhibit more risky, undisciplined, or "sloppy" behaviors (Zhang et al., 2014; Liu et al., 2015). For example, if a user exposes the organization's network to vulnerabilities present in chat services or peer-to-peer in an organization that disallows use of those services, the MSSP would report a network security policy violation. The number of such policy violations, as measured by the number of corresponding incident reports written by the MSSP, may be one of the many indicators of network misuse or mismanagement and point to an organization's greater vulnerability to intrusions. Although we hypothesize that violations are predictive of the number of intrusions, data collection and analysis of violations require potentially a comparable level of effort and subject matter expertise as required for collection of intrusion data. Therefore, in practice, this may be a roadblock to examining this variable as a predictor (see Section 5 for detailed analysis of results with and without this predictor included).

### 3.1.3 Cyber footprint

We characterize an organization's cyber footprint by three predictors (see Table 1): (i) the number of hosts; (ii) number of records on scholar.google.com (ROSG) per organization; and (iii) a categorical predictor of how closely each organization resembles an academic institution. Our study examines whether these features are predictive of intrusions (see Table 1 for additional details).

The number of hosts connected to a given organization's cyber sensor node provides the MSSP with insights into the vulnerability of the network (see Table 1). That is, intuitively, the higher the number of hosts in an organization's network, the more opportunities exist for penetration and exploitation of network and host-based vulnerabilities (i.e., cyber-attacks). However, the data for this covariate may only capture a fraction of the actual hosts and is an estimate. Nonetheless, to gain initial insights into whether this variable is predictive of the likelihood of a sensor node (or organization) to face intrusions, we also include it in the model predictors.

Through preliminary analyses of the number of intrusions and organization type, we conjecture that networks of organizations engaged in educational and research activities tend to have significantly greater number of intrusions. To test this hypothesis, we query scholar.google.com for a name of an organization, and use the number of records returned for that organization as a quantitative indicator, admittedly imperfect, of the organization's engagement in research activities. We further characterize, for each sensor, to what extent the corresponding organization is similar to an educational institution with the categorical predictor, similarity to educational institution behavior (SEIB). Specifically, strictly educational institutions were assigned the value 10; organizations with ROSG greater than 1,000 were assigned the value of 3; and for all other cases we assigned the value of 1. Together with the covariate ROSG, the categorical predictor SEIB (see Table 1) provides a subset of the characteristics that can be used to describe cyber footprint of each organization. One limitation of this predictor is that we are constrained by the data provided by the MSSP. However, there are additional potential predictors that may be analyzed, including additional types of organizations (e.g., military), country of origin or geolocation of hosts and DNS traffic.

## 3.2 Response variable

The response variable is the number of successful cyber intrusions per organization on its network reported to the MSSP within a multi-month time period. We examine the predictors (presented in Table 1 and described in Section 3.1) for their influence on the response. The predictors connect details about an organization's DNS activities with its cyber footprint and security posture, and we will show in Section 4 which of these variables best predict the response and to what extent.

Because our study necessarily hinges on empirical data, our findings and insights are limited to the scale of the data (e.g., number of observations, time period covered by observational data) and the types of predictors assessed.

# 4 Generalized linear models

We define four models to simulate the response data (see Section 3.2) in the most appropriate manner: (i) standard linear regression model, (ii) principal component (PC) regression model, (iii) Poisson GLM, and (iv) negative binomial (NB) GLM. The first of the models is a standard linear regression model. We show that this simple approach is inappropriate for the count data for two reasons: (i) using the ordinary least squares (OLS) method to estimate the coefficients of the standard linear regression model, the OLS-fitted response variable can take on negative and non-integer values[1] (Long, 1997); and (ii) the statistical analysis of the regression results exhibit (not covered in this paper for the sake of brevity) collinearity in the predictors with condition number of 27.4 which is greater than the recommended threshold value of 20. Consequently, the model has a low quality of fit. We resolve the collinearity in the predictors with PCA such that the dimension of the space of predictors was reduced from 12 to 8 principal components and approximately 99% of the variability in the data was still explained[2]. However, neither the standard linear nor the PC regression modeling approaches predicted the response well, in part, because the fitted values for both models can be negative (Long, 1997; Long and Freese, 2006).

In the following, we no longer consider standard linear or PC regression models, and instead we examine two GLMs commonly used for count data (e.g., in medical, epidemiological, and ecological literature): the Poisson and NB models (Frome, 1983; Lawless, 1987; McCullagh and Nelder, 1989; Hope et al., 2003; Woolhouse et al., 2015). These models extend the standard linear and PC regression models such that the response is assumed to be a random variable from the exponential distribution family with a nonlinear link function between the parameterization of the distribution of the response and a linear combination of the predictors. Note that standard linear regression can be viewed as a GLM, where only the mean parameter is being modeled, and the link function is simply the identity. Furthermore, the NB model provides an extension of the Poisson model that may better fit the response if the variance is greater than the mean response by definition (see Section 4.2). The following subsections define the two models that we evaluate for the best fit of the intrusions data.

## 4.1 Poisson GLM for count data

The probability distribution that we select for the number of intrusions, greatly impacts the fit of the GLM. Suppose that we have $m$ observations of the response (see Section 3.2), the number of intrusions, $y_1, \ldots, y_m$ that are Poisson random variables such that

$$y_i \sim \text{Pois}(\lambda_i), \quad \lambda_i = \exp\left(\sum_j x_{ij} \beta_j\right), \tag{1}$$

where $x_{ij}$ are the predictors for i=1, …, m and $j$ = 1, …, $n$, and $\beta_1, \ldots, \beta_j$ are the coefficients (or parameters) of the GLM (Greene, 2008). The mean and variance of the Poisson distribution are equal such that $E(y_i) = \text{var}(y_i) = \lambda_i$. In addition, we assume the response data, $y_i$ in (1), are conditionally independent given the predictors, and we associate the predictors with the response via the canonical log link function. By exponentiating the linear combination of predictors, we guarantee a positive mean response.

## 4.2 Negative binomial GLM for count data

The NB model is a generalization of the Poisson model presented in (1). For this model, the response variable is a Poisson-distributed random variable, $y \sim \text{Pois}(Z)$, where $Z$ is a random variable from the gamma distribution with $E(Z) = \alpha/\beta$ and $\text{var}(Z) = \alpha/\beta^2$. Let $\alpha = \beta = 1/\gamma$ and define the response values, the number of intrusions, as a random variable with the NB distribution in (2)

$$P\{Y = y\} = \frac{\Gamma(\alpha + y)\beta^\alpha \lambda^y}{y!\,\Gamma(\alpha)(\mu + \beta)^{y + \alpha}} \tag{2}$$

---

[1] For example, one organization in the data has an observed value of 149 intrusions, however, the mean OLS-fitted response value is –62.7.
[2] Prior to performing PCA on the predictors, we standardize the data such that it is mean-centered.

The NB probability distribution in (2) is also known as the gamma-Poisson mixture distribution (McCullagh and Nelder, 1989; Hilbe, 2007; Greene, 2008). In addition, we assume that there is the log link function for (2), where E(Z) = λ, var(Z) = λ(1 + γλ), and $\gamma$ is known as the heterogeneity parameter. Although other link functions are also commonly considered for the NB GLM, its canonical link function is the natural logarithm, and this version of the NB GLM is often referred to as the NB2 model (Greene, 2008). Understanding this model in terms of a mixture of the Poisson distributions in (1), where the parameters are distributed as per a gamma distribution, allows modeling of overdispersed data, where the variance exceeds the mean (Green, 2008). Note that if $\gamma = 0$, the variance equals the mean, and the NB2 and Poisson models are equivalent.

# 5 Results

We assess which of the two GLMs that we defined in Section 4–Poisson regression and NB2 models–provide the best fit for the response data, the number of intrusions per organization. We estimate the coefficients of the models in (1) and (2) with the maximum likelihood estimation method, iteratively re-weighted least squares (IRLS) (Frome, 1983) and implement the models in the Python programming language primarily using the Statsmodels (Seabold and Perktold, 2010) and Numpy (van der Walt et al., 2011) libraries.

In addition, we generalize the predictive performance of the models by implementing the jackknife (or leave-one out) procedure (Maloof et al., 2003; Abdi and Williams, 2010). With this method, we show how well the models perform on new observations by successively dropping each observation and fitting the regression models to the remaining set of observations. The model is then used to predict the left-out observation. To determine whether the response data follows the Poisson or NB2 probability distribution, we examine the models' regression results and goodness-of-fit statistics.

## 5.1 Comparison of the Poisson regression and NB2 models

In Table 2, we present the regression results for the Poisson regression and NB2 models, where the heterogeneity parameter $\gamma$ for the NB2 model is set to 0.01 to better model the overdispersion in the observed response data than the Poisson regression model. In Section 5.2, we assess further the impact of varying $\gamma$ for the NB2 model on the goodness-of-fit statistics. The statistical significance of the estimated coefficients differs between the two models (see Table 2), and varying $\gamma$ also impacts the regression results and significance of model parameters. Assessing the Poisson GLM, we find that 8 of the 14 coefficients have statistically significant estimates for the following predictors: domestic.com, domestic.net, domestic.org, foreign.net, hosts, violations, SEIB=3, and SEIB=10. In contrast, for the NB2 model with $\gamma = 0.01$, the predictors, domestic.com, domestic.org, and hosts are not statistically significant (see Table 2).

The Pearson residuals reveal additional information about the goodness-of-fit of the models and capture the deviation of each of the fitted values from the observed values, where the increase in aggregate measure Pearson $\chi^2$ indicates a declining model fit. The Pearson residuals, $p_i$, are defined in (3)

$$p_i = \frac{y_i - \hat{\mu}_i}{\sqrt{\text{var}(\hat{\mu}_i)}} \qquad (3)$$

where the $y_i$ represents the observed response, $\hat{\mu}_i$ is the expected value of the fitted response, and $\text{var}(\hat{\mu}_i)$ represent the variance of the fitted values for i = 0, 1, 2, ..., m. More simply, the Pearson residuals are the raw residuals scaled by the standard deviation of the response data, and they indicate a lack of fit when greater than 2 in absolute value (McCullagh and Nelder, 1989).

In Figure 1, we examine a plot of the observed and Poisson-fitted response values to visually inspect the goodness-of-fit of this model. In addition, we examine the Pearson residual dependence plot for the Poisson model and, as an additional discrepancy measure, the histogram for the standardized deviance residuals for these models (see Figure 1). A test for the quality of fit of the model based on the Pearson and deviance residuals is to determine whether the maximum of the residuals in absolute value are less than 2 (McCullagh and Nelder, 1989). Although there are 3 outliers in the predicted response variable for the
Poisson GLM as indicated by the associated Pearson residual values that are greater than 2 (for fitted response values less than 10), this model fits the response data very well for the response values greater than 10, where the Pearson residuals in absolute value are less than 2 (see Figure 1). In Figure 1, we also show the standardized deviance residuals for the Poisson regression model.

Table 2. Jack-knifed regression results for GLMs, where p-values below 0.05 are highlighted in bold, Poisson GLM has 26 residual DFs, and NB2 model has 25 residual DFs.

|  | Poisson regression GLM | | | NB2 with $\gamma = 0.01$ | | |
|---|---|---|---|---|---|---|
|  | Log-Likelihood: | −92.225 | | Log-Likelihood: | | −93.689 |
| Link Function: log | Deviance: | 78.006 | | Deviance: | | 75.911 |
| DF Model: 13 | Pearson $\chi^2$: | 89.2 | | Pearson $\chi^2$: | | 86.1 |
| Predictors | Coefficients | std err | P>\|z\| | Coefficients | std err | P>\|z\| |
| Intercept | −0.2442 | 0.277 | 0.377 | −0.256 | 0.512 | 0.617 |
| domestic.com | 2.97E-06 | 1.10E-06 | **0.007** | 3.01E-06 | 2.06E-06 | 0.143 |
| domestic.edu | −0.0003 | 0 | 0.108 | −0.0003 | 0 | 0.371 |
| domestic.gov | 0.0004 | 0 | 0.134 | 0.0004 | 0 | 0.411 |
| domestic.net | −7.52E-06 | 1.29E-06 | **0** | −7.45E-06 | 2.47E-06 | **0.003** |
| domestic.org | −5.42E-05 | 2.12E-05 | **0.011** | −5.43E-05 | 4.06E-05 | 0.181 |
| foreign.com | −6.08E-06 | 5.23E-06 | 0.246 | −6.00E-06 | 9.81E-06 | 0.541 |
| foreign.net | 7.00E-05 | 1.66E-05 | **0** | 6.97E-05 | 3.10E-05 | **0.024** |
| foreign.org | −4.47E-08 | 1.74E-07 | 0.798 | −4.40E-08 | 3.33E-07 | 0.895 |
| hosts | 2.55E-05 | 8.27E-06 | **0.002** | 2.54E-05 | 1.67E-05 | 0.128 |
| violations | 0.2126 | 0.038 | **0** | 0.2113 | 0.072 | **0.003** |
| SEIB=3 | 1.8739 | 0.253 | **0** | 1.8653 | 0.491 | **0** |
| SEIB=10 | 2.6341 | 0.584 | **0** | 2.6227 | 1.134 | **0.021** |
| ROSG | −4.44E-05 | 3.63E-05 | 0.221 | −4.40E-05 | 6.72E-05 | 0.513 |

The Pearson $\chi^2$ in Table 2 is a goodness-of-fit measure that aggregates the discrepancy between all observations and their respective predicted values (i.e., the Pearson residuals), where increasing values indicate a lack of model fit. The Pearson $\chi^2$ for the Poisson GLM is 89.2, and it is lower for the NB2 model (with $\gamma = 0.01$) at 86.1. This is one indication that the NB2 model is a better fit than the Poisson GLM. However, in Figure 2, we show that the NB2-fitted response values with $\gamma = 1.314$ do not fit the observed response data as closely (for two observed response values that are outliers) as the NB2 model with $\gamma = 0.01$. We will explore additional goodness-of-fit statistics in the subsequent subsections to compare with this assessment.

## 5.2 Overdispersion: Parametrizing the NB2 model

In this section, we examine the impact of varying the heterogeneity parameter $\gamma$ on the NB2 model's fit of the intrusions data by assessing additional goodness-of-fit measures, including the estimated dispersion parameter and the Bayesian Information Criteria (BIC). The estimated dispersion parameter is defined as the ratio of deviance $D(\hat{\beta})$ to the residual DFs, $\hat{\phi} = D(\hat{\beta})/m - n - 1$, where $D(\hat{\beta})$ is a measure residual variation that is the sum of the deviance residuals. In addition, deviance is defined as twice the negative logarithm of the likelihood ratio for the restricted model to the absolute or general model (with more parameters), $D(\hat{\beta}) = -2\{L(\hat{\beta}) - L(y)\}$, where $L(\mathbf{y})$ is defined as the log-likelihood function at $\hat{\mu}_i = y_i$, for $i = 1, \ldots, m$ (Frome, 1983). Furthermore, $D(\hat{\beta})$ is approximately an approximately $\chi^2$-distributed random variable with $m–n–1$ DFs.

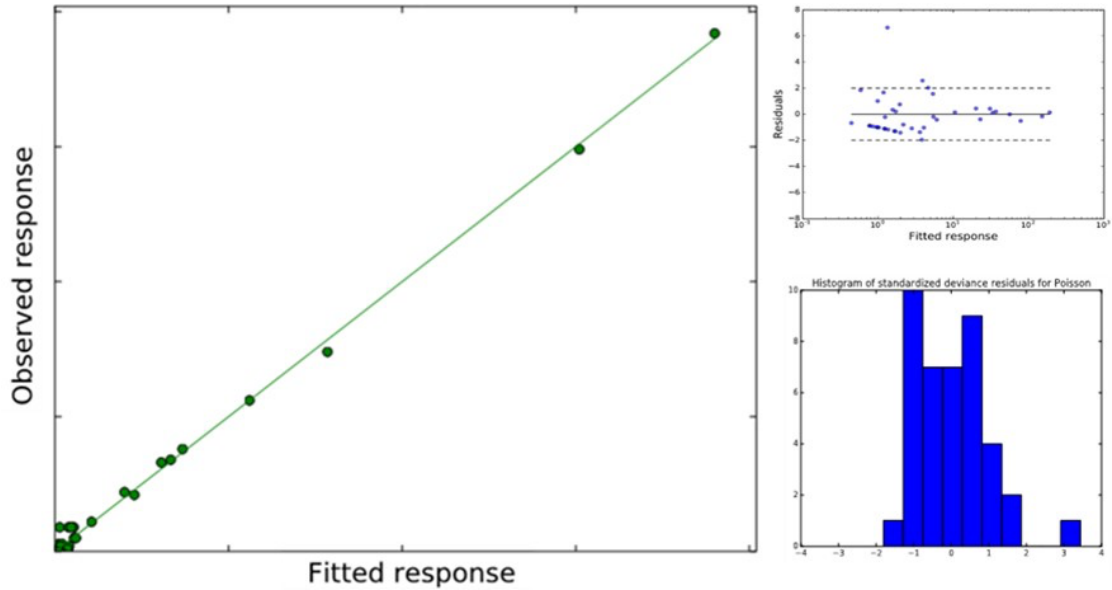

Figure 1. Nonnegative jackknifed Poisson regression model predictions for the number of intrusions on the x-axis (i.e., fitted response) and the observed number of intrusions on the y-axis are presented (left). On the upper right, a scatter plot of the Pearson residuals on the y-axis and the Poisson-fitted response on the x-axis (on a log-scale) combined with the dashed lines at $y = \pm 2$ are presented to highlight the outliers in the fitted values (i.e., above 2 in absolute value). A histogram of the standardized deviance residuals is also presented (lower right).

If the ratio of deviance to residual DFs exceeds 1, then there is overdispersion (see Tables 3 and 4), and the standard errors of the coefficients may be unreliable. Note that the NB2 model has 1 less residual DFs than the Poisson regression model due to its heterogeneity parameter, and as such, the Poisson regression model is a restricted version of the NB2 model. That is, if the heterogeneity parameter is set to 0, the two models are equal (see Section 4). In Tables 3 and 4, we present the estimated dispersion parameter and BIC for the Poisson regression and NB2 models, and we show that $\hat{\phi} = 3.0$ (i.e., the deviance is 3.0 times larger than its 26 DFs) for the Poisson model, an indication of overdispersion. Similarly, we find that $\hat{\phi} = 3.04$ for the NB2 model with $\gamma = 0.01$ (i.e., this model has 25 residual DFs, where the heterogeneity parameter is included in $n$).

We explore the impact of varying the heterogeneity parameter $\gamma$ between 0.01 and 1.5 on the overdispersion in the NB2 model, in part, by measuring the estimated dispersion parameter $\hat{\phi}$ and BIC for both models (see Tables 3 and 4), where BIC = $D(\hat{\beta}) - DF*\log(m)$, where recall that m is the number of observations, DF represents the number of residual DFs, and $D$ is the deviance statistic measured at the estimated model coefficients. Ideally to ensure a good quality model fit, $\hat{\phi}$ is close to 1.0, and the BIC measure is minimized across models to yield the best fit GLM for the response data. The Poisson regression and NB2 (with $\gamma = 0.01$) models have comparable BIC and $\hat{\phi}$ values (see Tables 3 and 4), but the NB2 model has a slightly better fit (i.e., lower BIC value). In addition, the maximum values of the Pearson residuals for the Poisson regression and NB2 (with heterogeneity parameter $\gamma = 0.01$) models are close at 6.64 and 6.51, respectively for a single fitted outlier in both models (see Figures 2 and 4). The NB2 model with $\gamma = 0.01$ also has a lower Pearson $\chi^2 = 86.1$ (as compared with the Pearson $\chi^2$ for Poisson GLM which is 89.2). However, the improved fit of the NB2 model with $\gamma = 0.01$ is not statistically significant, according to the log likelihood ratio test. Nonetheless, statistical analysis of the standardized deviance residuals for the Poisson regression and NB2 models provide a clear indication that the NB2 model ($\gamma = 0.01$) better fits the observed response data than the Poisson GLM. Although both the Pearson and standardized deviance residuals are useful for determining the adequacy of fit of the models, it is the deviance residuals that are minimized when the GLM is fit to the response data.

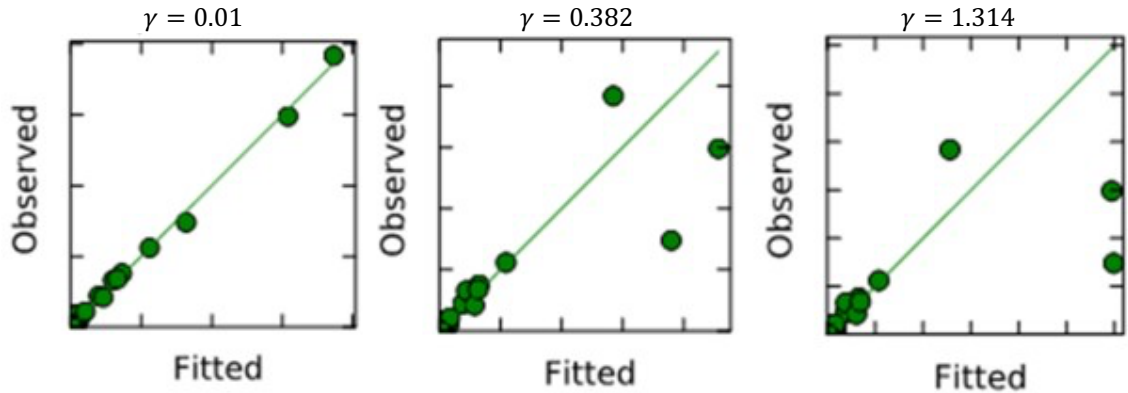

Figure 2. Nonnegative jackknifed predictions for the number of intrusions on the x-axes and observed number of intrusions on the y-axes are presented for the NB2 models with heterogeneity parameter $\gamma = 0.01, 0.382, 1.314$ from left to right.

As $\gamma$ increases for the NB2 model, the average BIC values decreases (an indication of improving fit). Based on the estimated dispersion parameter and BIC values alone in Table 4, we would select the NB2 model with $\gamma = 1.31$ as the best fit GLM. The estimated dispersion parameter $\hat{\phi} = 1.01$ and BIC = –56.23 for the NB2 model with $\gamma = 1.31$, compared with $\hat{\phi} = 3.04$ and BIC = –22.4 for the NB2 model with $\gamma = 0.01$ (see Table 4). Although this indicates overdispersion for $\gamma = 0.01$ and equidispersion for the model with $\gamma = 1.31$, the latter model does not fit two observed response values that are greater than 10 (see Figure 2). In fact, these 2 observations are outliers for the dataset in that the DNS traffic for these two sites are substantively greater than the other sites in our intrusions data. Although there is a lack of fit for these 2 outliers with the NB2 model for $\gamma > 0.01$ (see Figures 3 and 4), the NB2 model with $\gamma = 1.31$ fits the response values less than 10 best of the heterogeneity parameters considered (see Figure 3). The NB2 model with $\gamma = 1.31$ does not fit the outliers in the observations well but provides a great fit for the remaining observations. Given that the deviance residuals are minimized during the fit of the GLM, we examine these results in Figure 3 which show that the NB2 model with $\gamma = 0.01$ provides the best fit GLM for the observed response data. In addition, setting the heterogeneity parameter $\gamma$ to 0.01 provides the best model fit for the outliers in the response data compared with the fit of the NB2 model for other heterogeneity parameter values considered in our analysis (see Figure 2).

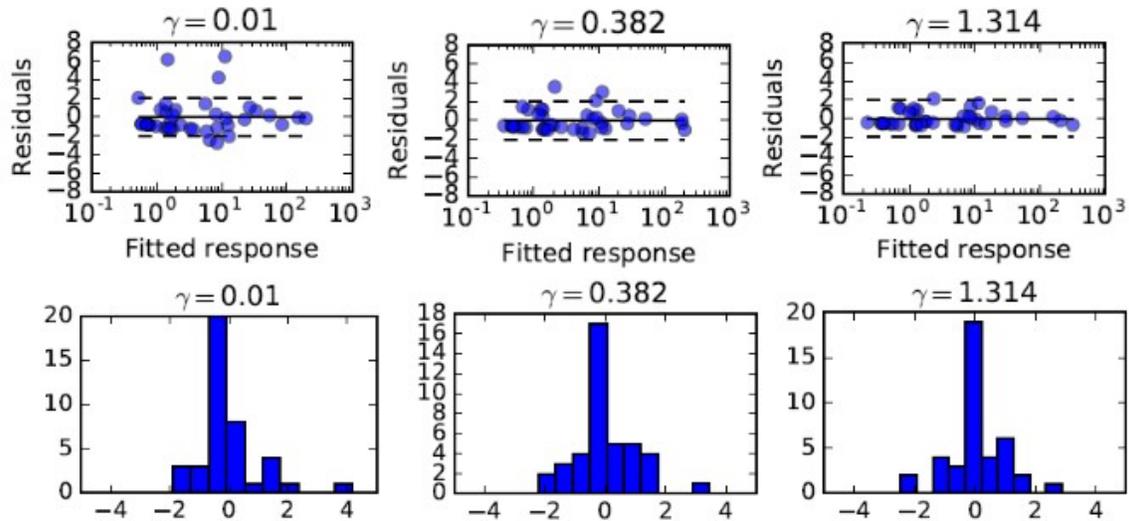

Figure 3. Pearson residuals dependence plots (top) and histograms for the standardized deviance residuals (bottom) are plotted for the NB2 models with the heterogeneity parameter $\gamma$ = 0.01, 0.382, 1.314. The Pearson residuals indicate the models consistently fit the observed response data better for larger fitted response values (greater than 10) than those fitted values closer to 0 (i.e., Pearson residuals in absolute value are less than 2 for fitted values greater than 10). Dashed lines are plotted at y = ±2 to represent lack of fit for residual values outside of this interval. The histogram of standardized deviance residuals for the model with $\gamma$ = 0.01 has a higher number of values less than 1 in absolute value (and near 0) than the other models, an indication that this is the best fit GLM.

## 5.3 Impacts of Omitting Selected Predictors

As we discussed in Section 3, an MSSP may have difficulty acquiring input data for the violations predictor in advance for a given organization. Hence, we explore further how well the models fit the observed response variable when restricted to exclude the violations predictor. Furthermore, in an effort to further assess the impact of cyber footprint on how well each of the models fit the response data, we also omit hosts, ROSG, and SEIB predictors from the full set of predictors (see Table 1). Specifically, we evaluate and compare the regression results for the models for five cases:
1. Omit the violations predictor
2. Omit the SEIB predictors
3. Omit the hosts predictor
4. Omit the ROSG predictor
5. Omit all cyber footprint predictors (i.e., SEIB, ROSG, and hosts)

Although we show the statistical significance of each of the estimated coefficients with the regression results in Table 2, we examine the results for the restricted models in the Cases presented above for several additional reasons, including: (i) SEIB is a qualitative measure that, in turn, may limit the reproducibility of the model fit to other datasets; (ii) the hosts predictor requires a system administrator for measurement or host-based security system (HBSS) which often only covers part of the organization's network; and (iii) the ROSG predictor may vary quite widely across organizations over time, and like (ii), the exact numbers may be difficult to measure depending on the complexity of an organization's networks and subnetworks. We compare the models with goodness-of-fit measures, including jackknifed BIC measures for the 5 cases, and we assess the models for overdispersion with the estimated dispersion parameter (see Tables 3 and 4). In Table 5, we present the jackknife predicted coefficients for the NB2 models for Cases 1-5 above, and their levels of statistical significance are indicated by (***), (**), and (*) for p-values less than 0.001, 0.01, and 0.05, respectively.

### 5.3.1 Case 1: Impact of Omitting the Violations Predictor

Because the violations coefficient is statistically significant for both GLMs considered (see Table 2), omitting it will impact the regression results for both models. Due to the difficulty that an MSSP may have in accurately assessing this predictor, we examine its impact on regression results if excluded from the NB2 model, and we find that only the estimated coefficients for the SEIB predictors, SEIB = 3 and SEIB=10, are statistically significant for this restricted model (see Table 5). Recall that SEIB is a qualitative variable with 3 categories: 1 is the reference category for our implementation, 3, and 10 (see Table 1). The estimated coefficients for the SEIB=3 and SEIB=10 predictors for the NB2 model (with $\gamma = 0.01$) are 2.5078 and 3.4782, respectively. This indicates that compared with the reference category of organizations (with SEIB=1), the expected log count for the group with SEIB=3, for example, increases by 2.5078 for the NB2 model. Furthermore, for both the Poisson and NB2 models, an omission of the violations predictor results in higher BIC values (e.g., Poisson GLM BIC = 9.7 compared with a BIC = –20.3 for the full Poisson GLM) implying a declining model fit (see Tables 3 and 4). Similarly, the restricted

NB2 model for Case 1 has a higher BIC value than the full NB2 model, where a lower BIC indicates that there is a better model fit (see Table 4). In addition, in Table 4, we present the estimated dispersion parameter for the restricted and full models for various heterogeneity parameter values to show the impact of restricting the models to this reduced set of predictors on overdispersion and model fit.

### 5.3.2 Case 2: Impact of Omitting the SEIB Predictor

In this section, we examine the impact of excluding the categorical predictor, SEIB, from the NB2 model and find that the estimated coefficients for the following predictors are statistically significant: domestic.com, domestic.edu, domestic.net, foreign.net, hosts, and violations (see Table 5). The omission of the categorical predictor, SEIB, results in a substantive change in the statistical significance of the estimated coefficients for the predictors, hosts, domestic.com, and domestic.edu, which are statistically insignificant in the absolute or general model (see Table 5 with NB2 model results for this case and $\gamma = 0.01$). Furthermore, the goodness-of-fit measures, log likelihood ratio test (p-value < 0.001), BIC, Pearson $\chi^2$, and deviance, indicate that the general model is a better fit of the observed response data than the restricted model without SEIB (see Tables 3 and 4). Omitting the SEIB predictors from the Poisson GLM also degrades the quality of fit for the Poisson model.

### 5.3.3 Case 3: Impact of Omitting the Hosts Predictor

Given the difficulty in acquiring data for the hosts predictors, we consider a restricted version of the full NB2 model ($\gamma = 0.01$), where we omit the hosts predictor from our regression analysis. We find that the estimated coefficient for domestic.org is now statistically significant; whereas, it is not for the general NB2 model. This is the only coefficient which changes its statistical significance with this restricted model fit. In addition, we show that the full model fits the intrusions data better than this case, where hosts are excluded (based on the log likelihood ratio test yielding a p-value = 0.006).

Table 3. Jackknifed BIC and estimated dispersion parameter, $\hat{\phi}$, for the 5 cases considered for the Poisson regression model.

| Poisson GLM | $\hat{\phi}$ | Avg. BIC | Std. dev. of BIC |
|---|---|---|---|
| Absolute model (all predictors) | 3.00 | –20.30 | 5.16 |
| Case 1 (omit violations) | 4.17 | 9.70 | 9.35 |
| Case 2 (omit SEIB) | 5.02 | 31.91 | 10.76 |
| Case 3 (omit ROSG) | 2.96 | –22.12 | 4.59 |
| Case 4 (omit hosts) | 3.24 | –14.57 | 5.11 |
| Case 5 (omit cyber footprint) | 6.59 | 81.68 | 12.23 |

### 5.3.4 Case 4: Impact of Omitting the ROSG Predictor

For all cases considered in Section 5.3, the estimated coefficient for the ROSG predictor is not statistically significant (see Table 5). Although removing ROSG from the set of predictors does not change the statistical significance of the NB2 model coefficients as compared with the full model, this restricted version of the model does improve the fit of the NB2 model for the response variable when compared to the absolute model (see Table 4 for lower BIC values for the restricted model than the absolute model). Furthermore, the log likelihood ratio test implies that the more complex model with the full set of predictors does not provide a statistically better fit than the simpler model with less parameters (i.e., p-value = 0.198). Given that we can predict the number of intrusions without assessing this predictor directly, this version of the NB2 model is a better choice for fitting the response data than the full model.

### 5.3.5 Case 5: Impact of Omitting All Cyber Footprint Predictors

Removing all cyber footprint predictors (i.e., hosts, SEIB, ROSG), substantively degrades the quality of fit for the NB2 model. We show the lack of fit for this restricted model in Tables 3 and 4, where the overall model goodness-of-fit measures, including BIC, deviance, and Pearson $\chi^2$ each indicate that the other cases considered yield a better fit for the response data. Furthermore, we find that the full model fits the observed response data significantly better than this restricted model by the log likelihood ratio test (with p-value < 0.001).

## 6 Discussion and Conclusions

We began this study by posing several research questions related to quantifying cyber risk: (i) whether a set of readily observable characteristics of an organization can be used to model the number of intrusions into its computer networks; (ii) what kind of model would be appropriate for the task; (iii) whether the model could be adequately accurate for practical purposes, such as informing the pricing and staffing of cyber defense services; and (iv) which, if any, of the initially conjectured predictors–cyber footprint, security posture, and DNS traffic–would be influential and applicable to the modeling task.

To answer these questions, we present four GLMs and analyze their regression results for adequacy of fit to the intrusions data, and we compare their predictions, given the interactions of the security posture, DNS traffic, and cyber footprint predictors considered: (i) standard linear regression model in which the response is a normally-distributed random variable, where the link function is the identity; (ii) PC regression that leverages PCA to reduce the dimensionality of the space of predictors and eliminate the collinearity impacting the validity of the statistical significance of estimated coefficients in (i) above; (iii) Poisson regression model, where the response are count data modeled as a Poisson random variable; and (iv) NB2 model in which the response is modeled as a NB2 random variable to fit the overdispersion in the observations.

Among our key findings is that one of these models – the generalization of the Poisson regression model to the NB2 model – predicts the response variable appreciably better than others. This is not too surprising given that the response variable represents overdispersed count data (Hilbe, 2007; Greene, 2008). Furthermore, we find that we can capture the overdispersion in the response by selecting a heterogeneity parameter ($\gamma = 1.31$) that improves the overall model goodness-of-fit measures considered (i.e., estimated dispersion parameter, $\hat{\phi} = 1.01$, and BIC $= -56.1$). However, when assessing the standardized deviance residuals, it is the NB2 model with $\gamma = 0.01$ that provides the best fit GLM. We also find that if $\gamma = 1.31$, this model does not fit two outliers in the observed response data well. Whereas, the NB2 model with $\gamma = 0.01$ provides a better model fit for the positive response values, by fitting the two outliers in the observations better (see Figure 2). Likely, there are two underlying distributions governing the response values: one distribution for the zeros (i.e., organization without any intrusions) and another for the positive number of intrusions. We plan to explore this further in future work.

We demonstrate that the intrusions data exhibit sufficient regularity (see for example the histogram of standardized deviance residuals in Figure 3), and the construction of a practically useful predictive model with the NB2 ($\gamma = 0.01$) model is feasible (see Tables 3, 4, and 5). To a large extent, these findings answer our first three research questions.

However, the fourth research question–which of the initially conjectured predictor variables should be included in the model– brought rather surprising findings. Several of the predictor variables that were recommended to us by subject matter experts

Table 4. Jackknifed BIC and estimated dispersion parameter, $\hat{\phi}$, for the 5 cases considered for the NB2 model, where the heterogeneity parameter is varied in this interval, $0.01 \leq \gamma \leq 1.5$.

| NB2 model | $\gamma$ | $\hat{\phi}$ | Avg. BIC | Std. dev. of BIC |
|---|---|---|---|---|
| Absolute model (all predictors) | 0.01 | 3.04 | −22.40 | 5.01 |
| | 0.20 | 2.14 | −33.09 | 11.49 |
| | 0.38 | 1.73 | −39.91 | 13.53 |
| | 0.57 | 1.49 | −44.82 | 14.50 |
| | 0.76 | 1.31 | −48.58 | 15.01 |
| | 0.94 | 1.19 | −51.60 | 15.29 |
| | 1.13 | 1.09 | −54.10 | 15.43 |
| | 1.31 | 1.01 | −56.23 | 15.49 |
| | 1.50 | 0.82 | −58.05 | 15.50 |
| Case 1 (omit violations) | 0.01 | 4.01 | 5.80 | 8.79 |
| | 1.31 | 1.18 | −47.35 | 23.60 |
| | 1.50 | 1.10 | -49.96 | 23.45 |
| Case 2 (omit SEIB) | 0.01 | 4.86 | 22.85 | 9.72 |
| | 1.31 | 1.08 | −49.99 | 30.58 |
| | 1.50 | 1.00 | −52.97 | 30.04 |
| Case 3 (omit ROSG) | 0.01 | 2.98 | −24.27 | 4.57 |
| | 1.31 | 0.97 | −59.29 | 15.93 |
| | 1.50 | 0.91 | −61.17 | 15.93 |
| Case 4 (omit hosts) | 0.01 | 3.21 | −18.21 | 4.81 |
| | 1.31 | 0.98 | −58.11 | 17.65 |
| | 1.50 | 0.92 | −60.09 | 17.56 |
| Case 5 (omit cyber footprint) | 0.01 | 6.22 | 65.12 | 10.86 |
| | 1.31 | 1.13 | −44.50 | 44.76 |
| | 1.50 | 1.05 | −48.52 | 43.71 |

(SMEs) turned out to be lacking in influence or even misleading. In particular, in Case 4, we omit ROSG from the original set of predictors, and also examine the influence of this omission (see Table 5). We find that the estimated coefficients for the ROSG predictor for both of the Poisson (see Table 2) and NB2 models and all cases considered are not statistically significant (see Table 5). SMEs felt that ROSG would be a significant predictor of intrusion frequency. We assess the log-likelihood ratio test comparing the full and restricted NB2 models yielding a p-value = 0.198 for the $\chi^2$-distribution in which the difference in residual DFs between the two models is 1. We conclude that ROSG alone is not a useful predictor of successful intrusions. However, the categorical predictor, SEIB, depends on ROSG, and SEIB has statistically significant influence on the response (see Table 5).

Table 5. Comparison of the estimated coefficients for the predictors of the NB2 regression model with $\gamma = 0.01$ for the five cases considered in Section 5.3. If the associated p-value is less than 0.001, 0.01, and 0.05, then (***), (**), and (*), respectively, are presented next to the predictor's coefficient to indicate its level of statistical significance.

|  | Full model | Case 1: no violations | Case 2: no SEIB | Case 3: no hosts | Case 4: no ROSG | Case 5: no cyber footprint |
|---|---|---|---|---|---|---|
| Log-likelihood: | −93.689 | −109.89 | −121.3 | −97.516 | −94.519 | −145.9 |
| Deviance: | 75.911 | 108.31 | 131.14 | 83.566 | 77.572 | 180.33 |
| Pearson $\chi^2$: | 86.1 | 109 | 139 | 85.1 | 82.4 | 213 |
| Residual DFs | 25 | 26 | 27 | 26 | 26 | 29 |
| Predictors | Estimated coefficients for each case | | | | | |
| Intercept | −0.256 | 0.6436 | −0.6922 | −0.1849 | −0.3677 | −0.3799 |
| domestic.com | 3.01E-06 | 6.82E-07 | 4.79E-06** | 3.70E-06 | 3.56E-06 | 6.03E-06*** |
| domestic.edu | −0.0003 | 0.0003 | −0.0007* | −0.0001 | −0.0003 | −0.0005 |
| domestic.gov | 0.0004 | 0.0002 | 0.0004 | 0.0002 | 0.0004 | 0.0002 |
| domestic.net | −7.45E-06** | −2.31E-06 | −7.57E-06** | −6.66E-06** | −7.37E-06** | −6.18E-06* |
| domestic.org | −5.43E-05 | −5.46E-05 | 2.90E-06 | −7.83E-05* | −6.30E-05 | −9.63E-06 |
| foreign.com | −6.00E-06 | 1.13E-05 | −1.47E-05 | −1.71E-06 | −4.19E-06 | −1.40E-05 |
| foreign.net | 6.97E-05* | −7.77E-06 | 8.25E-05** | 5.24E-05 | 6.22E-05* | 7.03E-05* |
| foreign.org | −4.40E-08 | −1.42E-07 | −3.03E-07 | −7.14E-08 | −2.48E-09 | −6.09E-07 |
| Hosts | 2.54E-05 | 8.30E-06 | 5.42E-05** |  | 2.17E-05 |  |
| violations | 0.2113** |  | 0.2936*** | 0.1736** | 0.2079** | 0.254** |
| seib3 | 1.8653*** | 2.5078*** |  | 2.0495*** | 1.7218*** |  |
| seib10 | 2.6227* | 3.4782** |  | 2.6987* | 2.2938* |  |
| ROSG | −4.40E-05 | −3.05E-05 | 2.48E-05 | −1.44E-05 |  |  |

Yet another variable that the SMEs expected to be influential–the number of hosts within an organization's network–also turns out to be a less significant predictor for the NB2 GLMs than hypothesized by SMEs (see Table 5). We base this conclusion on the fact that the estimated coefficients for the hosts predictor is statistically significant for only Case 2 of the NB2 model, where the restricted model omits the SEIB predictor (see Table 5). However, excluding the number of hosts slightly degrades the model's quality of fit (i.e., the log likelihood ratio test yields a p-value = 0.006).

On the other hand, we show that the number of violations is a strong predictor of number of intrusions (see Tables 2 and 4). In Case 1, we exclude the policy violations predictor and examine the influence of its omission on our predictions and the statistical analyses for overall model goodness of fit and the distribution of residuals. We find that excluding this predictor from the models considerably impacts the drivers of intrusions (see Table 5) such that only the SEIB predictors remain statistically significant in this case. Thus, policy violations predictor is crucial for constructing a successful model, and cannot be omitted.

Specifically, the models show that the violations predictor consistently has significant impact on the number of intrusions (see Tables 2 and 4) for each of the cases considered in Section 5.3. Using the NB2 model with heterogeneity parameter $\gamma = 0.01$, we predict that the number of intrusions increases with each additional violation, the expected response is $e^{0.2113}$ times larger (see Table 2). This finding is rather intuitive. Indeed, if users lack in discipline or knowledge to comply with organizational cyber hygiene policies, and if the organization is unable or unwilling to enforce its own policies, it is easy to expect that the organization's cyber defenses are poor, leading to more frequent intrusions.

Less intuitive is our finding that the variables, numbers of domestic.net and foreign.net domain name visits, are strong predictors of intrusions. Although it is not entirely clear why this should be the case, a possible explanation, or at least a partial support to this finding, could be found in the corpus of domain data assessed by Spamhaus, an international nonprofit organization that tracks spam and related cyber threats (e.g., phishing, malware, botnets). In this data, the .net domain (see https://www.spamhaus.org/statistics/tlds/) has the highest "badness score," where badness is defined as

$$\frac{\text{number of bad domains observed}}{\text{total number of domains observed}} \log(\text{number of bad domains observed}).$$

With this scoring, the Spamhaus Project data shows that .com and .org domains have badness scores significantly lower than those of .net in Table 6. Furthermore, .mil, .edu, and .gov domains are all associated with a 0 badness score (see Table 6). There are tighter controls on the granting of .mil, .edu, and .gov domains. To obtain one of these domains, proof that the request organization is a US military, educational, or government entity must be provided.

Table 6. Spamhaus Project badness scoring for top TLDs (https://www.spamhaus.org/statistics/tlds/).

| TLDs | % of bad domains | Badness Score |
| --- | --- | --- |
| .net | 14.9% | 1.66 |
| .com | 5.3% | 0.66 |
| .edu | 0.0% | 0.00 |
| .org | 5.4% | 0.50 |
| .gov | 0.0% | 0.00 |
| .mil | 0.0% | 0.00 |

Whereas, for .com, .org, and .net domains, no proof is required that the organization is commercial, a non-profit (as was originally intended for .org), or involved in networking technologies (e.g., Internet Service Provider, networking hardware manufacturer). Therefore, it could be conjectured that the .net TLD is more likely to serve as a carrier of malicious infrastructure.

In the future, we plan to assess the quality of fit of the zero-inflated NB2 regression model for the intrusions data. Furthermore, we plan to explore additional data sources that are specifically related to network security posture, such as other network mismanagement and misuse symptoms (see Section 3). An additional future direction of our research on predicting the number of successful cyber intrusions is to explore ways of using the predictive models as inputs to pricing models for cyber security and insurance services. Another avenue for future work is to inform cyber risk dynamics, by modeling how the number of intrusions per organization varies over time as a function of the temporal variations in relevant organizational features. To this end, we will model the temporal dynamics of cyber intrusion with additional data; such as the inter-arrival time of cyber intrusions, additional TLDs, and more features of network mismanagement symptoms.